\documentclass[onecolumn,amsmath,amssymb,aps,pra,longbibliography,12pt]{revtex4-1}
\usepackage{url}
\usepackage[breaklinks=true,colorlinks=true, linkcolor=blue,urlcolor=blue,anchorcolor=blue,citecolor=blue,bookmarksnumbered]{hyperref}
\usepackage{graphicx}
\usepackage{dcolumn}
\usepackage{bm}

\begin{document}
\title{Rydberg Superatom Interface for Topological Microwave-to-Optical Photon Conversion in Fock-State Lattices}	
\author{Pei-Yao~Song$^{1}$}\author{Jin-Lei~Wu$^{1}$} \email[]{jlwu517@zzu.edu.cn}\author{Weibin~Li$^{2}$}\email[]{weibin.li@nottingham.ac.uk}\author{Shi-Lei~Su$^{1,3}$}\email[]{slsu@zzu.edu.cn}
	\affiliation{$^{1}$School of Physics, Zhengzhou University, Zhengzhou 450001, China\\
	$^{2}$School of Physics and Astronomy, and Centre for the Mathematics and Theoretical Physics of Quantum Non-equilibrium Systems, The University of Nottingham, Nottingham NG7 2RD, United Kingdom\\
	$^{3}$Institute of Quantum Materials and Physics, Henan Academy of Science, Henan 450046, China} 
	
\begin{abstract}
Microwave-to-optical conversion (MTOC) of single photons plays a pivotal role in bridging quantum devices across different frequency domains, but faces challenges in maintaining efficiency and robustness against fluctuations and dissipation in hybrid quantum systems. Here, we propose a topologically protected MTOC scheme mediated by a Rydberg superatom to address these limitations. By constructing cross-linked Fock-state lattices (FSLs) through a dual-mode Jaynes-Cummings (JC) architecture, we map the effective hybrid system onto an extended Su-Schrieffer-Heeger~(SSH) model with tunable hopping rates. Photon-number--dependent property of hopping rates triggers a topological phase transition in the extended SSH chain, converting the defect mode into a topological channel that directionally pumps photons between microwave and optical cavities. This mechanism leverages Rydberg blockade-enhanced photon-superatom couplings to establish a robust energy transfer channel, achieving high-efficiency photon conversion under realistic decoherence. Our theoretical framework demonstrates how topological protection synergizes with Rydberg-mediated light-matter interactions to realize a robust quantum transducer, providing a scalable platform for noise-resilient quantum networks and frequency-multiplexed quantum interfaces.
\end{abstract}
\maketitle
	
\section{INTRODUCTION}\label{sec1}
Topological edge pumping emerges as a robust energy transfer mechanism governed by the system geometric phase, fundamentally connecting bulk topological invariants with edge-state transport dynamics~\cite{Serra-Garcia2018,Kraus2012,Jurgensen2021Nature,Cheng2022NC,Citro2023NRP,Song2024SciAdv,Wu2024NC}. This mechanism has been extensively studied in quantum and classical light-based systems, demonstrating ability to achieve robust lightwave transport between distant nodes~\cite{Lu2014,Kiczynski2022,doi:10.1126/science.ado8192}. Recently, the concept of Fock-state lattices (FSLs) has been proposed as a synthetic-dimensional platform engineered in quantum optical systems~\cite{caihan2020NSR,PhysRevLett.116.220502}, where discrete Fock states (photon-number states) map to lattice sites, and coherent field couplings between adjacent states emulate ``hopping" in synthetic space. This framework enables the simulation of topological phenomena in reduced physical dimensions, such as classical photonic and acoustic waveguides~\cite{PhysRevA.110.063529,PhysRevA.107.033501,yuan2021APLphotonics,PhysRevApplied.21.064068} and classical phononic and microwave resonators~\cite{Yang2024,PhysRevB.109.125123,PhysRevLett.134.116601}. Beyond these classical physical platforms, while FSLs have been successfully implemented in quantum regimes~\cite{doi:10.1126/science.ade6219,PhysRevLett.134.070601}, their deployments in quantum information process remain largely unexplored.
The integration of topological protection mechanisms with quantum information processing protocols in FSL architectures thus constitutes an open frontier, particularly for applications requiring noise-resilient photon-number-state manipulation, for example microwave-to-optical conversion~(MTOC) of single photons.

On the one hand, the exploration of efficient MTOC of single photons is a key challenge in quantum information processing, bridging the gap between microwave-based quantum computing platforms and optical quantum networks~\cite{Tu2022,Petrosyan_2019,Borowka2024,Liu21OE}, which is of great importance for applications in hybrid quantum networks, long-distance quantum communication, and quantum memory~\cite{PhysRevLett.125.260502,Zhong2021,Jiang2023,vanThiel2025,Chen2023,Rochman2023}. However, due to the large frequency mismatch between microwave and optical modes, traditional coupling methods often suffer from low conversion efficiency and susceptibility to environmental noise~\cite{Forsch2020,Jiang2020,PhysRevLett.127.040503,Meesala2024}. On the other hand, the implementation of MTOC by using advantages of {\it topological} quantum state transfer~(QST) remains unexplored, although topological transports in robust quantum state transfer have been widely investigated in quantum platforms~\cite{PhysRevA.98.032323,PhysRevLett.105.255302,Jotzu2014,PhysRevLett.123.063001,PhysRevA.110.043510}. Therefore, developing a quantum interface capable of mediating topologically protected MTOC presents a critical scientific frontier. To this end, we consider here the Rydberg superatom, formed by a mesoscopic ensemble of Rydberg atoms in the regime of Rydberg blockade~\cite{PhysRevLett.87.037901,superatom2024su,PhysRevResearch.2.043339}, to serve as an interface for topological MTOC of single photons. Uniquely, the collectively enhanced coupling between the superatom emitter and the photon modes can be much larger than the decay rates, leading to a strong coupling regime of an effective Jaynes-Cummings (JC) model~\cite{PhysRevA.82.053832}. These position Rydberg superatoms as a promising interface candidate for implementing MTOC.

In this work, we extend the theoretical framework of topological defect states in FSLs~\cite{caihan2020NSR} to a Rydberg superatom-mediated MTOC system. By designing a hybrid dual-mode JC model among a MW cavity, a Rydberg superatom spin, and an optical cavity, we project the Fock states of the hybrid quantum system onto an extended one-dimensional (1D) Su-Schrieffer-Heeger (SSH) model. Specifically, we design a fast and robust topological state transfer by cross-linking two FSLs, and the temporal modulation of photon-number--depended hopping rates renders the cross-linked FSL to undergo a series of topological phase transitions, transforming the defect mode of the FLS into a topologically protected channel that facilitates the transfer of photons from the microwave cavity to the optical cavity. Our results demonstrate that the proposed MTOC device exhibits strong robustness against moderate perturbations in coupling strengths and cavity decay, simultaneously possessing scalability. This work provides a new perspective to achieve efficient conversion between microwave and optical photons, filling the gap in topological state transmission for quantum information processing in microwave-optical quantum networks.

\vspace*{-5pt}

\begin{figure*}
	\centering
	\includegraphics[width=\linewidth]{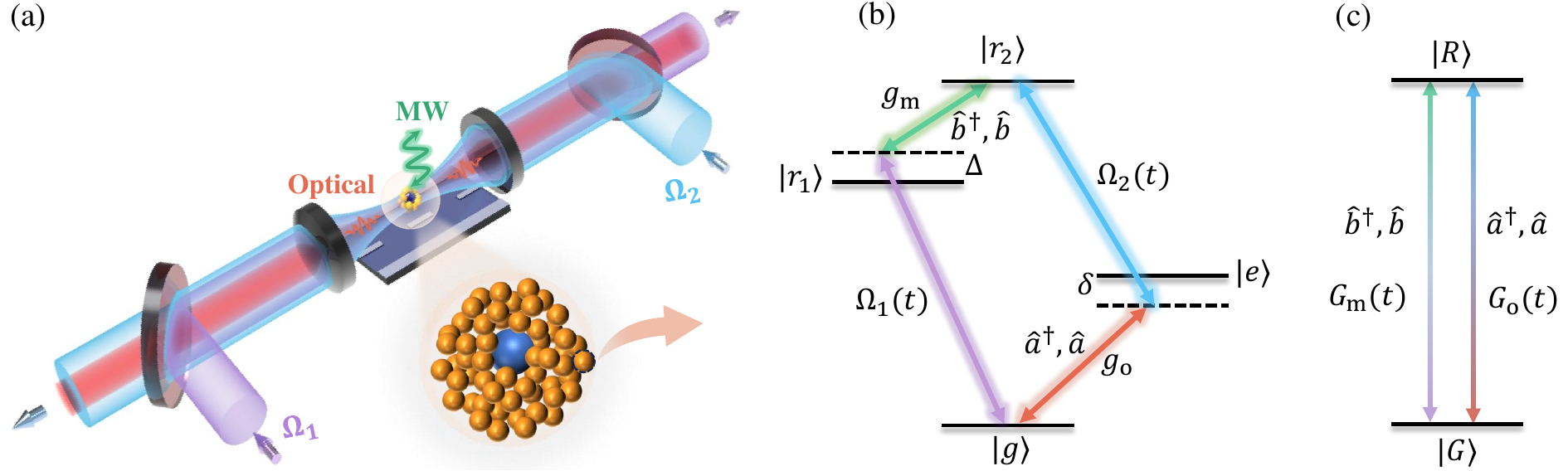}
	\caption{Schematic diagram of our MTOC system. (a)~Structure of the optical-microwave superatom interface. The atoms in ensemble~(superatom) are coupled simultaneously to an optical cavity and a superconducting coplanar waveguide (SCW) resonator. (b)~Sketch of the single-atom excitation process from the ground state $|g\rangle$ to higher-energy Rydberg state $|r_2\rangle$ through four-wave mixing. (c)~Microwave-optical dual-mode JC model with a superatom spin.}\label{fig1}
\end{figure*}
\section{HYBRID DUAL-MODE JC MODEL AND TOPOLOGICAL FSL}\label{sec2}
Our quantum transducer scheme is schematically depicted in Fig.~\ref{fig1}(a), considering a cold ensemble of $^{85}\rm{Rb}$ atoms confined in an optical microtrap, coupled simultaneously to an optical cavity and a microwave cavity, i.e., superconducting coplanar waveguide (SCW) resonator. $^{85}\rm{Rb}$ is chosen over the more typical $^{87}\rm{Rb}$ due to its higher natural isotopic abundance, which facilitates efficient loading from a Rb dispenser into the magneto-optical trap and ensures a sufficient number of atoms for the formation of stable Rydberg atoms~\cite{Kumar2023}. In the MTOC configuration, atoms in the ensemble are off-resonantly driven from the ground $|g\rangle$ ($5S_{1/2}$) to the Rydberg state $|r_1\rangle$ ($35P_{1/2}$) using a 297 nm UV laser with the time-dependent Rabi frequency $\Omega_1(t)$~\cite{Li2016,Thoumany09}, as shown in Fig.~\ref{fig1}(b). Due to the detuning $\Delta$, an atom can resonantly absorb a 297 nm UV photon only if it then also absorbs a signal MW photon (occupying mode $b$), resulting in a collective excitation to the Rydberg state $|r_2\rangle$ ($36S_{1/2}$) with the coupling strength $g_{\rm m}$. Meanwhile, the atom is coupled to a 481 nm laser driving a transition from the Rydberg state $|r_2\rangle$ to the intermediate state $|e\rangle$ with the Rabi frequency $\Omega_2(t)$ and detuning $\delta$. The atomic transition $|r_2\rangle\leftrightarrow|e\rangle$ is also connected to the ground state by emitting a 780 nm signal photon into the optical cavity mode $a$ with a coupling strength $g_{\rm o}$~\cite{PhysRevLett.121.123603}, thereby completing the transduction cycle. The driving light elucidates the process of employing standard four-wave mixing for MTOC~\cite{PhysRevA.110.052608}.

The single-atom Hamiltonian in the interaction picture is described by~($\hbar=1$)
\begin{eqnarray}
\hat{H}_{\rm s} &=& \frac{\Omega_1(t)}{2}|g\rangle\langle r_1|e^{i\Delta t}+g_{\rm m}|r_2\rangle\langle r_1|\hat{b} e^{i\Delta t} +\frac{\Omega_2(t)}{2}|e\rangle\langle r_2| e^{i\delta t}+g_{\rm o}|e\rangle\langle g|\hat{a}e^{i\delta t} +\rm{H.c.}
\label{eq1}
\end{eqnarray}
where ${\hat a}^{\dagger}$ (${\hat b}^{\dagger}$) and $\hat{a}$ ($\hat{b}$) are the creation and annihilation operators of the optical cavity (SCW resonator) mode, respectively. When the large detuning conditions $\Delta\gg|\Omega_1|,g_{\rm m}$ and $\delta\gg|\Omega_2|,g_{\rm o}$ are assumed, we obtain an effective single-atom Hamiltonian by adiabatically eliminating the two short-lifetime intermediate states~\cite{JJ2007}
\begin{eqnarray}
\hat{H}_{\rm eff}&=&\frac{g_{\rm m}\Omega_1(t)}{2\Delta}|r_2\rangle \langle g|\hat{b} +\frac{g_{\rm o}\Omega_2(t)}{2\delta}|r_2\rangle \langle g|\hat{a}+\rm{H.c.},
\label{eq2}
\end{eqnarray}
for which the Stark shift terms have been omitted because they can be canceled by inducing an opposite Stark shift with additional lasers in experiment~\cite{sciadv.aau5999,Wu2021,PhysRevA.98.032306}.
These transitions mediate the couplings of a superatom spin to the microwave and the optical modes, simultaneously, as shown in Fig.~\ref{fig1}(c), under the condition where the atomic ensemble is confined within the Rydberg blockade radius (thereby forming a Rydberg superatom); see Appendix A~\ref{appendixA} for details. The system can be expressed by a hybrid microwave-optical dual-mode JC model for the superatomic spin, described by Hamiltonian
\begin{equation}
\hat{H}_{\rm D}=G_{\rm m}(t)|R\rangle\langle G|\hat{b}+G_{\rm o}(t)|R\rangle\langle G|\hat{a}+\rm{H.c}.,
\label{eq3}
\end{equation}
with effective superatom-enhanced coupling strengths $G_{\rm m}(t)=\sqrt{N_a}g_{\rm m}\Omega_1(t)/2\Delta$ and $G_{\rm o}(t)=\sqrt{N_a}g_{\rm o}\Omega_2(t)/2\delta$~($N_a$ is the number of atoms in the ensemble). In our MTOC protocol, it is assumed that time-dependent Rabi frequencies are modulated as $\Omega_1(t)=\Omega_{\rm 1m}\sin(\pi t/2T)$ and $\Omega_2(t)=\Omega_{\rm 2m}\cos(\pi t/2T)$, with $T$ being the total modulation duration corresponding to the topological pumping period.  

Considering the $N$-excitation Hilbert subspace of the hybrid dual-mode JC model, a 1D modulated photonic FSL containing $2N + 1$ sites will be constructed, where the coupling strengths between adjacent sites vary with the square root of the index $j$ $(j = 1, 2,\cdots, N)$~\cite{caihan2020NSR,PhysRevLett.116.220502}. The resulting FSL sites govern photon dynamics with the following Hamiltonian
\begin{eqnarray}\label{eq4}
\hat{H}(t)&=&\sum_{j=1}^{N}u_j(t)|2j-1\rangle\langle2j| +v_j(t)|2j\rangle\langle2j+1|+\rm{H.c.},
\end{eqnarray}
where $u_j(t)=G_{\rm m}(t)\sqrt{N-j+1}$ and $v_j(t)=G_{\rm o}(t)\sqrt{j}$ are the hopping rates between two neighboring sites. The hopping parameters $G_{\rm m}(t)$ and $G_{\rm o}(t)$ are temporally modulated according to $G_{\rm m}(t) = g \sin(\pi t / 2T)$ and $G_{\rm o}(t) = g \cos(\pi t / 2T)$, where $g$ represents the maximum strength. Concretely, we simulate 1D FSLs by systematically mapping from the Fock states $|n_1,n_2,G/R\rangle$ of the dual-mode JC model with total excitation number $N$, where the non-negative integer $n_{1,2}$ is the photon number in the optical (microwave) modes, satisfying $m+n_1+n_2=N$ ($m=0$ for the superatomic ground state $|G\rangle$, while $m=1$ for the excited state $|R\rangle$).

\begin{figure}
	\centering
	\includegraphics[width=0.6\linewidth]{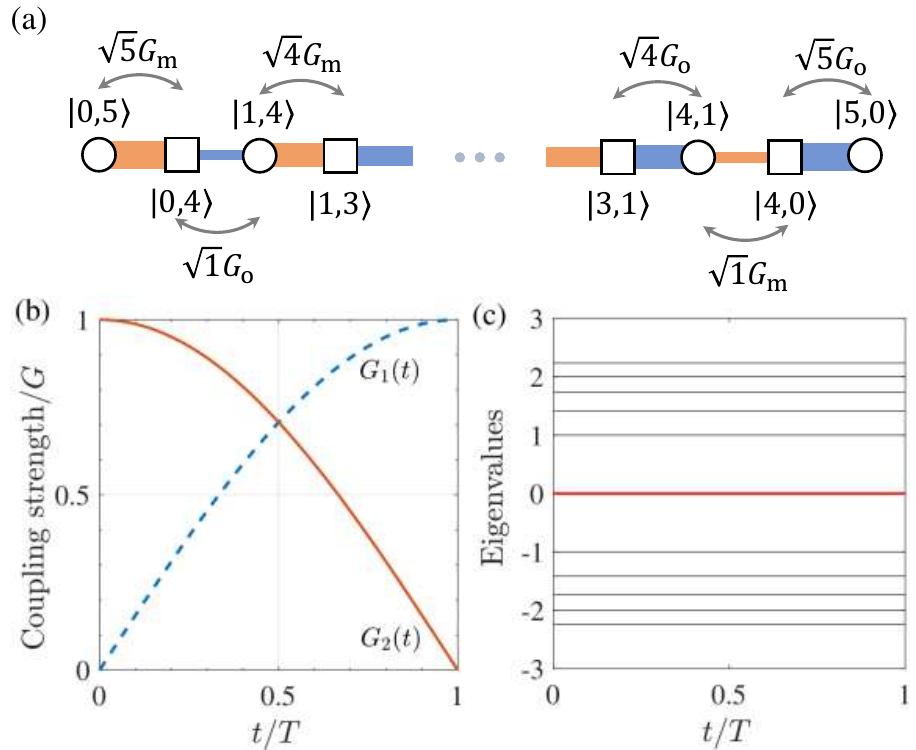}
	\caption{(a)~Schematic illustration of a 1D FSL with 11 sites with $t$-dependent staggered-square-root coupling strengths, in which the empty circles (squares) denote the state $|G\rangle$ ($|R\rangle$) of superatom. (b) Temporally modulated coupling parameters $G_{\rm m}(t) = g \sin(\pi t/2T)$ and $G_{\rm o}(t) = g \cos(\pi t/2T)$. (c) Time-dependent spectrum of Hamiltonian in Eq.~(\ref{eq4}).}\label{fig2}
\end{figure}
We take a Hilbert subspace of $N=5$ as an example in this work, and all photons are initially prepared in the SCW resonator, while the superatom is prepared in the ground state, represented by the state $|0,5,G\rangle$, which can be flexibly implemented in experiments~\cite{doi:10.1126/science.ade6219,Hofheinz2008}. These Fock states exactly constitute a 1D tight-binding lattice, where $|n_1,n_2,R\rangle$ is coupled to $|n_1,n_2+1,G\rangle$ and $|n_1+1,n_2,G\rangle$ with coupling strength $\sqrt{n_2+1}G_{\rm m}$ and $\sqrt{n_1+1}G_{\rm o}$, respectively, as shown in Fig.~\ref{fig2}(a). Specifically, the square-root-scaling coupling strengths between the FSL sites are owing to the nature of the bosonic annihilation operator, which breaks the translational invariance of the system~\cite{Lohse2016,PhysRevLett.107.103601}. In this extended SSH lattice with inhomogeneous hopping strengths, the topological invariant (winding number) can be characterized as $W=1-{\rm sign}(\prod_{j}u_j^{2}-\prod_{j}\nu_j^{2})$~\cite{PhysRevB.83.155429,RevModPhys.82.3045}.
The eigenstates of such a 1D chain can be analytically solved by introducing collective bright and dark modes, defined through their respective annihilation operators
\begin{eqnarray}
\hat{c}_1=1/{\sqrt{G_{\rm m}^2+G_{\rm o}^2}}(G_{\rm m}\hat{b}+G_{\rm o}\hat{a}),\cr\cr
\hat{c}_2=1/{\sqrt{G_{\rm m}^2+G_{\rm o}^2}}(G_{\rm o}\hat{b}-G_{\rm m}\hat{a}).
\label{eq5}
\end{eqnarray}
The Hamiltonian in Eq. (\ref{eq3}) can thus be reduced to
\begin{eqnarray}
\hat{c}_1=\sqrt{G_{\rm m}^2+G_{\rm o}^2} |R\rangle \langle G|\hat{c}_1+\rm{H.c.},
\label{eq6}
\end{eqnarray}
in which only the bright mode is coupled with the superatom. In this case, the eigenstates are easily obtained
\begin{eqnarray}
|\phi_{\pm j}\rangle&=&\frac{1}{\sqrt{2}}(|j,N-j;G\rangle_c \pm|j-1,N-j;R\rangle_c),
\label{eq7}
\end{eqnarray}
with corresponding eigenergies $E_{\pm j}=\pm\sqrt{j(G_{\rm m}^2+G_{\rm o}^2)}$.
In addition, there exists a zero-energy state due to the decoupled nonlocal mode $c_2$
\begin{eqnarray}
|\phi_0\rangle &=&\frac{(c_2^{\dagger})^{N}}{\sqrt{N!}}|0_{\rm o}\rangle\otimes |0_{\rm m}\rangle\otimes|G\rangle\nonumber\\
&=&\sum_{n_1,n_2}\sqrt{\frac{N!}{n_1!n_2!}}(\frac{G_{\rm o}}{G})^{n_2}(-\frac{G_{\rm m}}{G})^{n_1} |n_1,n_2;G\rangle,
\end{eqnarray}
where $|0_{\rm o}\rangle\otimes |0_{\rm m}\rangle$ is the vacuum state of the both cavities. Correspondingly, in the FLS chain the zero-energy state can be expressed by
\begin{equation}
|\phi_0\rangle =\sum_{k=0}^N\sqrt{\frac{N!}{(N-j)!j!}}(\frac{G_{\rm o}}{G})^{N-j}(-\frac{G_{\rm m}}{G})^j|2j+1\rangle,
\end{equation}
which only occupies odd-number FSL sites, centered where the two adjacent coupling strengths are equal. It can be seen that with modulated coupling strengths $G_{\rm m}(t)$ and $G_{\rm o}(t)$ in Fig.~\ref{fig2}(b), the energy spectrum does not change with time and presents a fully flat band result shown in Fig.~\ref{fig2}(c). This fully flat band property is thoroughly different from other standard SSH-based transmission schemes where exists an extremely narrow band gap during the transmission process, resulting in a decrease in adiabatic evolution efficiency~\cite{PhysRevA.98.012331,PhysRevB.99.155150,Cheng2022,PhysRevA.105.L061502,PhysRevLett.126.054301}.

Each intracell coupling in a dimer is stronger than the right-adjacent intercell coupling between two dimers on the left of the extended SSH model in Eq.~\eqref{eq4} (winding number $W = 0$), indicating a trivial topological phase, while on the right, we get a nontrivial topological phase with the winding number $W = 1$. In this situation, the zero-energy state acts as a defect in which the excitation profile is localized at the interface between two sublattices with different phases~\cite{science2018defect,Dong2017}. The density distribution of this zero-energy state is $|\phi_0|^2 \propto (G_{\rm o}/G)^{2(N-j)} (G_{\rm m}/G)^{2j}/(N-j)! j!$. Using Stirling's approximation, the maximum of the distribution is found from~\cite{mermin1984stirling}
\begin{eqnarray}
\frac{\partial}{\partial j} \ln |\phi_0|^2 \propto \ln \left( \frac{G_{\rm m}^2}{G_{\rm o}^2} \frac{N-j}{j} \right) = 0,    
\end{eqnarray}
yielding $G_{\rm m}/\sqrt{N-j} = G_{\rm o}/\sqrt{j}$. Thus, the zero-energy state is centered at $j = N/[1 + (G_{\rm m}/G_{\rm o})^2]$. This position depends critically on $G_{\rm m}/G_{\rm o}$, suggesting that exotic topological pumping can be achieved via dynamic moving of the defect states by modulating $G_{\rm m}$ and $G_{\rm o}$.
\begin{figure*}
	\centering
	\includegraphics[width=\linewidth]{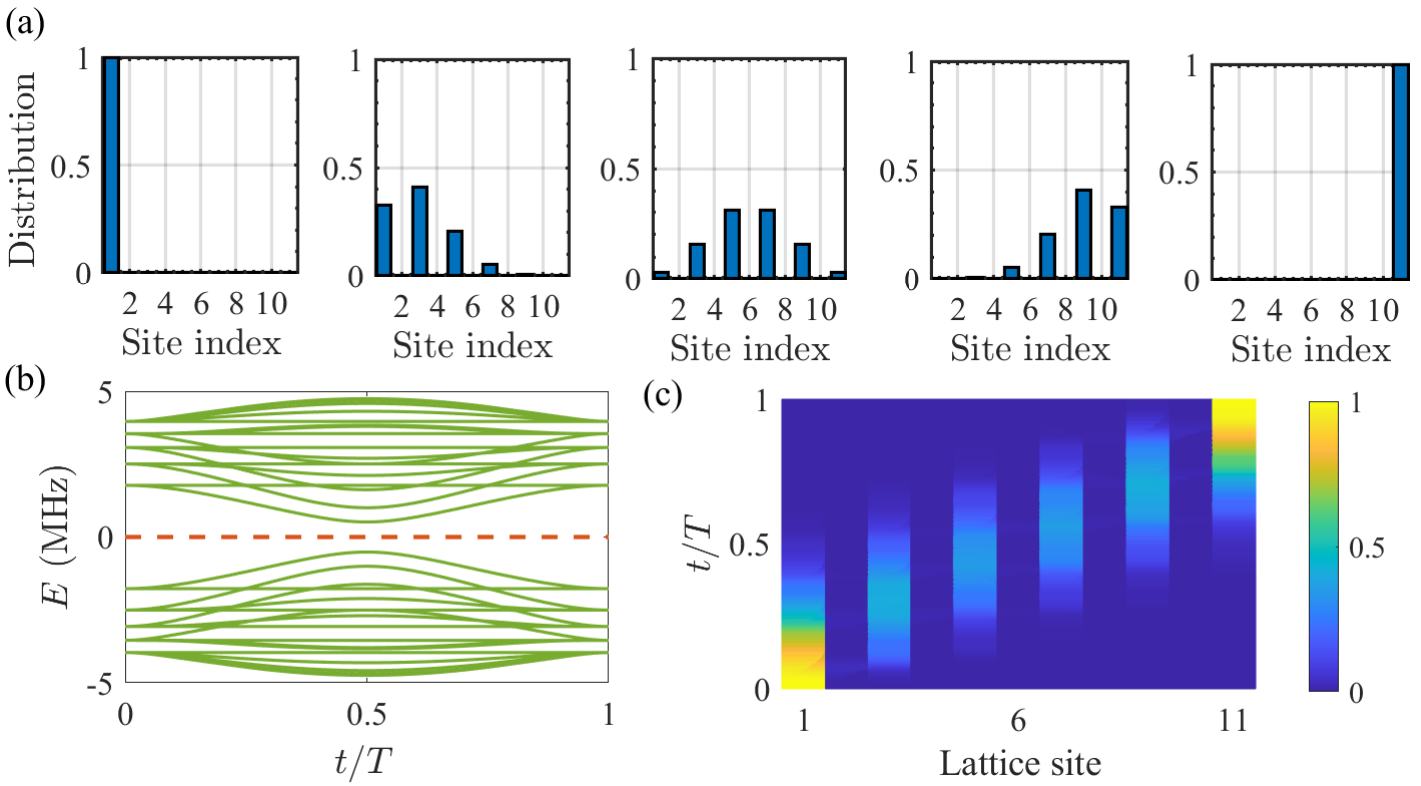}
	\caption{(a) Excitation distribution of the zero-energy mode~(defect state) in the FSL with $N=5$ at different coupling strength ratio $r=G_{\rm m}/G_{\rm o}$. From left to right, $r=\inf$~($G_{\rm o}=0$), $r=1/2$, $r=1$, $r=2$, and $r=0$~($G_{\rm m}=0$). (b) Energy spectrum of the dual-mode JC model in Eq.~(\ref{eq3}), with photon operators being numerically set up to 5-excitation dimension. (c) Time-dependent distribution on 11 FSL sites for zero-energy state of the 5-excitation dual-mode JC model.}\label{fig3}
\end{figure*}

\section{TOPOLOGICAL PUMPING OF THE DEFECT STATE IN FSL}\label{sec3}
During achieving the efficient topological photon transports for the MTOC scheme through the defect state between two topological phases, the evolution of the quantum hybrid system is protected by a constant energy gap $|E_{\pm 1}-E_0|=g\sqrt N$, with the position of interface in the FSL depending on the coupling strength $G_{\rm m}$ and $G_{\rm o}$. Figure~\ref{fig3}(a) demonstrates the distribution of the zero-energy defect state for different values of $G_{\rm m}/G_{\rm o}$. When the ratio $G_{\rm o}/G_{\rm m} >\sqrt{N}$ or $<1/\sqrt{N}$, the lattice fully enters one of the two distinct topological phases. When $G_{\rm m}=0$ and $G_{\rm o}=g$, the zero-energy state is strictly localized at the left edge of the chain. Alternatively, it becomes localized at the right end when $G_{\rm m}=g$ and $G_{\rm o}=0$. For intermediate coupling conditions, where $1/\sqrt{N}<G_{\rm o}/G_{\rm m}<\sqrt{N}$, the 1D chain exhibits a transition between two topologically distinct phases of the SSH model. This exotic feature makes the FSL lattice capable of realizing topological pumping of photons between the SCW resonator and the optical cavity through the dynamic transfer of defect states.

For parameters in a realistic physical system, we consider the superatom composed of 600 Rydberg atoms~\cite{Kumar2023}, adopting the coupling strengths $\sqrt{N_a}g_{\rm m}=2\pi\times4.5$ MHz and $\sqrt{N_a}g_{\rm o}=2\pi \times 5$ MHz, which can be realized using the most advanced experimental techniques available by choosing the single-atom optical cavity coupling $g_{\rm o}=2\pi\times206$ kHz and the single-atom microwave cavity coupling $g_{\rm m}=2\pi\times182$ kHz~\cite{Kumar2023}. Other parameters are $g=2\pi\times0.282$ MHz, $\Omega_{\rm 1m}=\sqrt{N_a}g_{\rm m}$, $\Omega_{\rm 2m}=\sqrt{N_a}g_{\rm o}$, $\Delta=2\pi\times70.5$ MHz, and $\delta=2\pi\times88.6$ MHz. For visualization, we plot the energy spectrum in Fig.~\ref{fig3}(b) based on the superatom-based dual-mode JC Hamiltonian Eq.~\eqref{eq3}. Regardless of the time-varying spectral lines that are decoupled to the 5-excitation subspace of the dual-mode JC model, it is identical to the spectrum in Fig.~\ref{fig2}(b) of the extended SSH model in FSL. When expanding the zero-energy state in the 5-excitation subspace, its time-dependent density distribution on each FSL site can be obtained, as shown in Fig.~\ref{fig3}(c). For the adiabatic transmission along the zero energy state, one can observe that the initial state only occupying the leftmost lattice site evolves to the final state locating the rightmost site, that is, all photons in the optical cavity.

\begin{figure*}
	\centering
	\includegraphics[width=\linewidth]{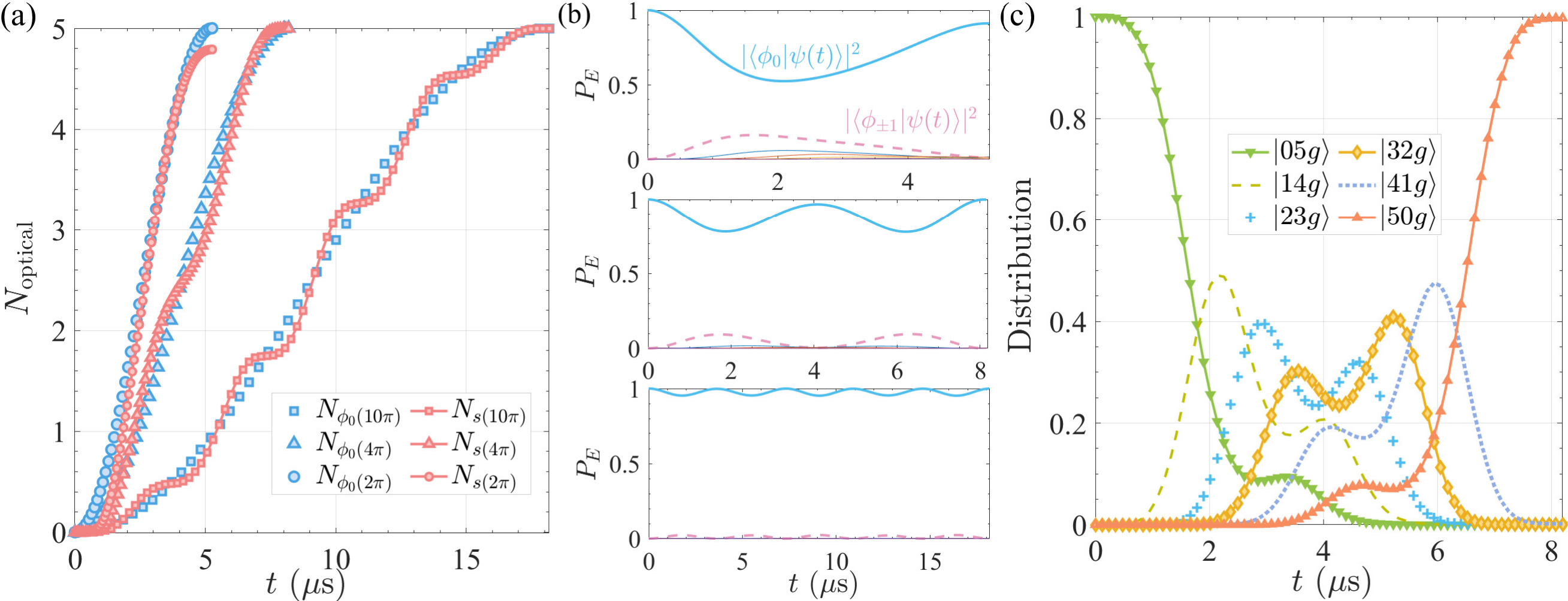}
	\caption{(a) Numerical simulation of optical photon number $N_{\rm optical}$ evolving with time. The Blue circles, triangles, and squares represent data derived by the zero-energy state $|\phi_0\rangle$ under the  effective areas of $2\pi$, $4\pi$, and $10\pi$, respectively. The corresponding red solid lines are calculated by solving the Schr\"odinger equation with the dual-mode JC model. (b) Population of the eigenstates $|\langle\phi_n|\psi(t)\rangle|^2$ with $|\psi(t)\rangle$ being the solution of Schr"odinger equation. The populations of the zero energy state $|\phi_0\rangle$ and the adjacent bulk states $|\phi_{\pm1}\rangle$ are marked with blue solid lines and pink dashed lines, respectively. From top to bottom, the evolution durations are $T_1=5.26$~$\rm \mu s$, $T_2=8.2$~$\rm \mu s$, and $T_3=18.18$~$\rm \mu s$, respectively. (c) Time-dependent distributions of the states located on odd-numbered FSL sites with $N=5$, where three elements in kets correspond to the photon number in the optical cavity, the photon number in the SCW resonator, and the superatom state, respectively.} \label{fig4}
\end{figure*}

The implementation of the MTOC scheme relies on evaluating the number of optical photons in the system at the end of the evolution, which signifies the conversion of photon frequency from the microwave to the optical regime.  Accordingly, to assess the  transmission performance, we initially plot the transmission dynamics in Fig.~\ref{fig4} without considering any dissipation effects. Specifically, in Fig.~\ref{fig4}(a) we simulate the optical photon number $N_{\rm optical}$ varying with time $t$ based on two methods: Numerically calculating the zero-energy state~($N_{\phi_0}$, blue marks); Solving Schr\"odinger equation of the dual-mode JC model~($N_{s}$, red solid-line marks). As expected, the time-varying optical photon numbers based on the dual-mode JC model exhibit a high degree of concordance relative to their counterparts derived by the the zero-energy state.

In Fig.~\ref{fig4}(a), we have picked three critical values of pumping duration according to the area theorem $A(N,T)\equiv gT=2n\pi$ with $n$ being a positive integer~\cite{PhysRevB.109.125123}, which determines a series of critical values of $T$ so as to ensure the system finally turning back to the zero-energy state after population oscillation between $|\phi_0\rangle$ and $|\phi_{\pm1}\rangle$. We illustrate three typical topological transfers with pulse areas $A = 2\pi, 4\pi$, and $10\pi$ in Fig.~\ref{fig4}(a), respectively. It should be noted that the actual critical time exhibits a slight deviation from the theoretically expected ones, which is supposed to be caused by ignoring population leakages to higher-order eigenstates due to nonadiabatic transitions. When the duration is shorter, the effect of nonadiabatic transitions will be more significant. Therefore, for $A=2\pi$ with $T= 5.26~{\rm \mu s}$, non-negligible population leakages to higher-order eigenstates lead to obvious deviation of final optical photon number from the ideal value 5. On the contrary, for $A=2n\pi$ with $n>1$, the population oscillations of the zero energy state can return well back to the zero-energy state after $n$ oscillation cycles. Taking the given three topological pumping durations $T =5.26~{\rm \mu s}$ ($A=2\pi$), $T =8.2~{\rm \mu s}$ ($A=4\pi$) and $T=18.18~{\rm \mu s}$ ($A=10\pi$) as examples, the oscillations of zero-energy population unambiguously indicates that the system state converges to the zero-energy mode at the end of topological pumping; see Fig.~\ref{fig4}(b).

As for the latter two higher-fidelity cases, the population evolution with $A=10\pi$ shows slighter oscillations since the coherent population exchange between $|\phi_{0}\rangle$ and $|\phi_{\pm1}\rangle$ is minimized by suppressing the nonadiabatic transitions, which is demonstrated by $P_{0}(t) =|\langle\phi_{0}(t)|\psi(t)\rangle|^2$ as well as the calculated population evolution of $P_{\pm1}(t) =|\langle\phi_{\pm1}(t)|\psi(t)\rangle|^2$, with $|\psi(t)\rangle$ being the solution of the Schr\"odinger equation with respect to the hybrid dual-mode JC Hamiltonian. Compared to the final optical photon number $N_{\rm optical}=4.994$ for $A=4\pi$ ($T=8.2~\rm \mu s$), the effective area $A=10\pi$ is almost approaching adiabatic evolution with $N_{\rm optical}=4.999$, which sacrifices evolution time but only achieves slight fidelity improvement. Prolonged evolution will unavoidably lead to intensified decoherence effects. Therefore, we intend to achieve high-fidelity photon conversion in nonadiabatic conditions with a shorter period by choosing the trade-off pumping duration $T=8.2~{\rm \mu s}$. Figure~\ref{fig4}(c) illustrates the evolving distribution of the FSL sites, which enables the population transfer of quantum states from the full microwave photons ($|05G\rangle$) to the full optical photons ($|50G\rangle$) within approximately $8\textsc{~}\rm \mu s$. It is worth noting that, due to the absence of population on even-numbered lattice sites, we naturally ignore the corresponding five states involving the superatom excitation.

This rapid transmission of photons from the microwave resonator to the optical cavity highlights the capability to achieve topological QST under nonadiabatic conditions. In contrast to traditional QST schemes, which necessitate extended transmission times to meet the adiabatic evolution and consequently result in accumulation of unavoidable decoherence and noise into practical quantum systems, our scheme can significantly improve state transfer performance. Furthermore, the excittaion distribution in our scheme is confined exclusively to odd-numbered FSL sites, with exceedingly low occupancy of the superatom excitation, indicating a pronounced suppression of spontaneous emission from atoms. Overall, the nonadiabatic transmission scheme, combined with the long-lived coherence of superatoms and the distinctive feature of the little population at even-numbered FLS sites, endows the MTOC mechanism in our scheme with high photon number scalability and photon transport robustness.

\section{ROBUSTNESS AND SCALABILITY}\label{sec4}
Distinct from traditional adiabatic topological transmission where the evolution fully follows the zero-energy mode, here the QST for MTOC is implemented with a small mixture of nonadiabatic tunnelings breaking the gap protection. Therefore, there is a natural question of whether the high-ﬁdelity QST scheme is still robust to decoherence and disorder. The performance of MTOC schemes is inevitably influenced by negative factors, such as decoherence, coupling strength stability, and the scalability of the system, which may degrade transmission efficiency and practical applicability.

\begin{figure}\centering
	\includegraphics[width=\linewidth]{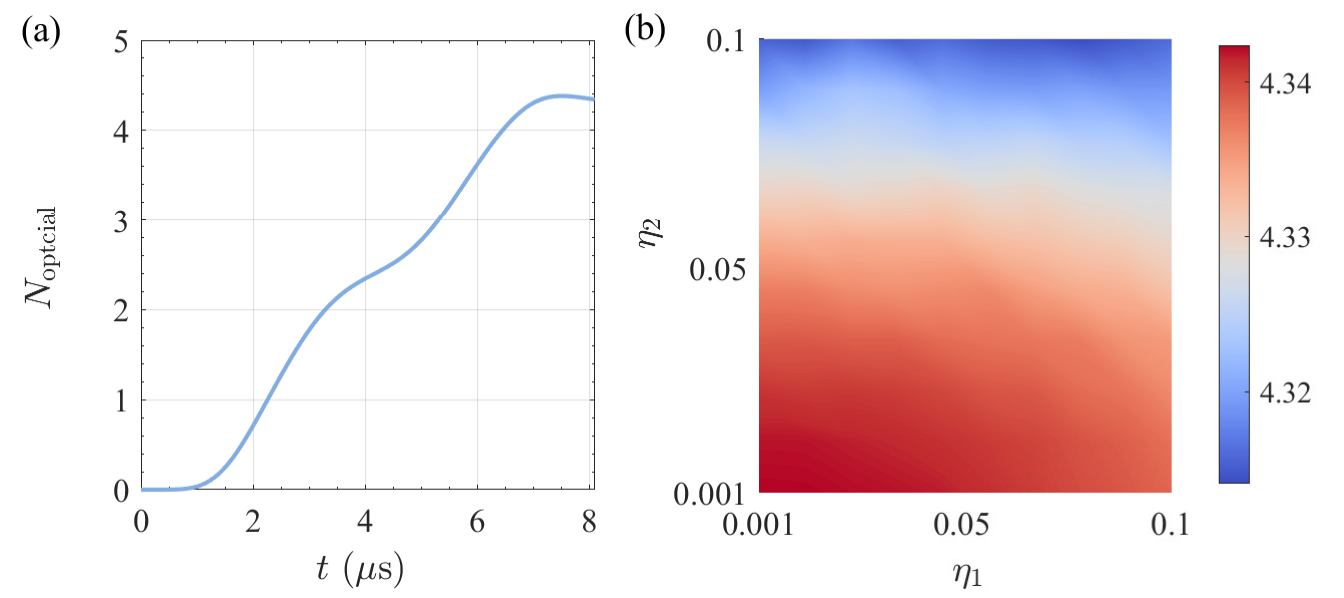}
	\caption{(a) Time-dependent average optical photon number $N_{\rm optical}$ with $T=8.2~{\rm \mu s}$, the decay rate of SCW resonator $\kappa_{\rm m}=2\pi\times2$ kHz and optical cavity $\kappa_{\rm o}=2\pi\times3.4$ kHz. (b) $N_{\rm optical}$ (indicated by the color bar) as a function of $\eta_{1(2)}$ for the coupling strength of superatom and the microwave resonator (optical cavity) $\sqrt{N_a}g_{\rm m(o)}(1+\epsilon_{1(2)})$, with $\epsilon_{1(2)}$ are randomly picked from the range $[-\eta_{1(2)},\eta_{1(2)}]$. Each datum in (b) denotes the mean of 1 001 results.}\label{fig5}
\end{figure}
In a realistic hybrid quantum system for our scheme, the main decoherence influence over the transmission fidelity is the decay of both cavities as well as the spontaneous decay of the excited superatom. To this end, we consider that the evolution of our MTOC system is governed by the Lindblad master equation
\begin{equation}
\dot{\rho}=i[\rho,\hat{H}_{\rm D}]+\hat{\mathcal{L}}(\hat{\sigma}_0)+\kappa_{\rm o}\hat{\mathcal{L}}(\hat{a})+\kappa_{\rm m}\hat{\mathcal{L}}(\hat{b}),
\end{equation}
where $\rho$ is the density matrix of the quantum system. $\hat{\mathcal{L}}(\hat{\sigma}_0)=(2\hat{\sigma}_0\rho\hat{\sigma}^{\dagger}_0-\hat{\sigma}^{\dagger}_0\hat{\sigma}_j\rho-\rho\hat{\sigma}^{\dagger}_0\hat{\sigma}_0)/2$ is the Lindbladian operator with $\hat{\sigma}_0=\Gamma_0(|G\rangle\langle R|)$, where $\Gamma_0$ describes the spontaneous emission rates from the superatom Rydberg state $|R\rangle$ to the ground state $|G\rangle$. $\hat{\mathcal{L}}(\hat{a})=(2\hat{a}\rho\hat{a}^{\dagger}-\hat{a}^{\dagger}\hat{a}\rho-\rho\hat{a}^{\dagger}\hat{a})/2$ and $\hat{\mathcal{L}}(\hat{b})=(2\hat{b}\rho\hat{b}^{\dagger}-\hat{b}^{\dagger}\hat{b}\rho-\rho\hat{b}^{\dagger}\hat{b})/2$ are the Lindbladian operators related to cavity decay, with the decay rates of the optical cavity $\kappa_{\rm o}$ and microwave cavity $\kappa_{\rm m}$, respectively.

We begin to focus on the evolution of optical photon number $N_{\rm optical}$ during the conversion process under the dissipative dynamics. The lifetime of the superatom Rydberg state $|R\rangle$ is about 44 $\rm \mu s$ at the temperature set to several millikelvin in the refrigerator~\cite{PhysRevX.14.031055,entanglement_science}, with the corresponding decay rate $\Gamma_0=2\pi\times3.6$ kHz~\cite{SIBALIC2017319}. For the decay rate of the SCW resonator and the optical cavity, we adopt the experimentally achievable parameters, in which $\kappa_{\rm m}=2\pi\times2$ kHz~\cite{Pirkkalainen2015} and $\kappa_{\rm o}=2\pi\times3.4$ kHz~\cite{doi:10.1126/science.abo3382}. As illustrated in Fig.~\ref{fig5} (a), by numerically solving the master equation, photons initially occupying the microwave cavity, corresponding to zero optical photons, and are are then gradually transferred into the optical cavity over time. At the end of topological pumping, approximately 4.4 optical photons can be successfully generated. Before $t=7.5~\rm \mu s$, the unitary dynamics dominates the evolution, the average optical photon number reaches around 4.5. In contrast, when the dissipative effect is sufficiently accumulated after $t=7.5~\rm \mu s$, the optical photon number subsequently declines slightly, highlighting the trade-off between the two dynamic processes, and the relatively slight degradation of the optical photon number indicates the robustness of our MTOC scheme to the system decoherence.

To verify the robustness of our topological MTOC scheme against disorders, here we consider control errors as disorders arising from fluctuations in the coupling strengths $\sqrt{N_a}g_{\rm m}$ and $\sqrt{N_a}g_{\rm o}$, in addition to the system dissipation. We assume that the coupling strengths of the two cavities with the superatom in the presence of fluctuations become $\sqrt{N_a}g_{\rm m}(1+\epsilon_1)$ and $\sqrt{N_a}g_{\rm o}(1+\epsilon_2)$, respectively, where $\epsilon_{1,2}$ represent the unwanted errors. We randomly pick sufficient~(1 001) pairs of data within the range $\epsilon_{1(2)}\in[-\eta_{1(2)},\eta_{1(2)}]$ to calculate the mean of the optical photon numbers. The results are plotted in Fig.~\ref{fig5}(b), which indicated that as $\eta_1$ and $\eta_2$ vary from 0.001 to 0.1, the average optical photon numbers fluctuate with the maximum magnitude only around 0.02. Specifically, even with a pair of $\sim10\%$ relative errors, the average optical photon number can still exceed 4.3, highlighting the giant robustness of our MTOC scheme against disorder due to the topological protection nature.

\begin{figure}
	\centering
	\includegraphics[width=0.6\linewidth]{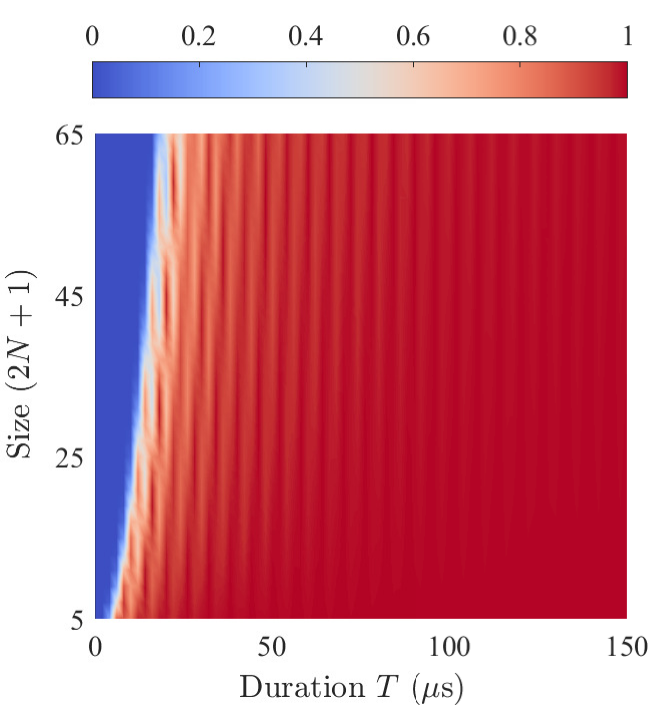}
	\caption{Final transfer fidelity for topological pumping of photons in MTOC scheme, calculated by solving the master equation with varying excitation number and evolution duration.}\label{fig6}
\end{figure}
Finally, we demonstrate the scalability for a highly efficient MTOC scheme. Although topological protection offers robustness against disorder, the practical implementation of MTOC in large-scale quantum networks requires that the system architecture be extendable without compromising performance. Therefore, it is crucial to evaluate whether our proposed scheme can be generalized to larger systems while maintaining topological advantages and operational faithfulness. To this end, we analyze the dependence of MTOC state transfer dynamics on the FSL size~(excitation number $N$) of the quantum system. In Fig.~\ref{fig6}, we calculate the state transfer fidelity in our topological MTOC scheme with varying FSL size $5\leq2N+1\leq65$ and pumping duration $0<T\leq150~{\rm \mu s}$, for which the transfer fidelity is defined as the population of the last site in the FSL, i.e., all of $N$ photons located in the optical cavity. It is shown that these fast topological passages with a small admixture of nonadiabatic oscillations are according to the pulse area of the three-level approximation as mentioned above; see apparent fringes in Fig.~\ref{fig6}. High-ﬁdelity topological MTOC exists stably where the pulse area is $4\pi$, $6\pi$, $8\pi$ and beyond, which determine the corresponding optimum of duration of accelerating MTOC without lowering the transmission fidelity when increasing $N$. Therefore, even in large-scale situations, high-fidelity nonadiabatic topological MTOC can be achieved. For pumping duration $T>100~{\rm \mu s}$, the nonadiabatic effect becomes gradually insignificant, when the transmission process of 1D chains tends towards an adiabatic regime.

\vspace*{6pt}

\section{DISCUSSION}\label{sec5}
In this work, we present a topological framework for achieving robust MTOC through Rydberg superatom-mediated quantum state transfer. By establishing a dual-mode JC model and implementing Fock-state mapping in an extended synthetic lattice, we have successfully engineered topologically protected energy transfer channels within a 1D SSH configuration. This approach harnesses the synergistic advantages of Rydberg blockade-induced collective superatom coupling to enable a fast and robust topological edge state transfer with the participation of nonadiabatic transitions.

The results of our simulations highlight several critical advantages of the proposed MTOC scheme. First, the system achieves high conversion efficiency, with a predicted optical photon number of about 4.5 from 5 microwave photons during a relatively short time of approximately 8 $\rm \mu s$, significantly outperforming traditional adiabatic topological QST methods that are prone to decoherence over longer timescales. Furthermore, the topological protection inherent in the SSH model ensures robustness against moderate perturbations in coupling strengths and cavity decay rates, and highly efficient photon conversion scheme can still be performed under $10\%$ error levels and experimentally accessible dissipation conditions. Moreover, the scalability of the scheme is validated across larger FSL sizes, maintaining high fidelity at critical transmission durations determined by the area theorem, thus offering a practical pathway for integration into large-scale quantum networks.

The underlying physics stems from locating hybrid microwave-superatom-optical tripartite states at odd-numbered FSL sites, which effectively suppresses atomic spontaneous emission through the zero-energy mode in the synthetic dimension. This topological engineering not only bridges the microwave-optical frequency divide but establishes a new paradigm for quantum information transduction between superconducting circuits and photonic devices. Our results could advance quantum hybrid interfacing technology in three aspects: First, the proposed nonadiabatic control scheme enables speed-fidelity optimization critical for noise-tolerant quantum information. Second, the inherent error resilience addresses key challenges in scalable quantum network construction. Third, the synthetic dimension approach may open new possibilities for implementing higher-order topological phenomena in quantum optical systems. These achievements could position Rydberg-mediated topological conversion as a cornerstone technology for future modular quantum architectures and cross-platform quantum internet development.

\appendix

\setcounter{equation}{0} 

\section*{Appendix A: Collective Coupling Enhancement Effect of Superatom}\label{appendixA}
\renewcommand{\theequation}{A\arabic{equation}}
Based on the Hamiltonian of a single atom in Eq.~(\ref{eq2}) in the main text, the Hamiltonian of the Rydberg atomic ensemble can be expressed as~\cite{superatom2024su}
\begin{eqnarray}
\hat{H}_I&=&\sum_{i=1}^{N_a}[\frac{g_{\rm m}\Omega_1(t)}{2\Delta}|r_{2i}\rangle\langle {g}_i|\hat{b}+\frac{g_{\rm o}\Omega_2(t)}{2\delta}|r_{2i}\rangle\langle {g}_i|\hat{a}]+{\rm H.c.}-\frac{1}{2}\sum_{i\neq k}\frac{C_6}{R_{ik}^6}|r_{2i}r_{2k}\rangle \langle r_{2i}r_{2k}|,
\end{eqnarray}
where a negative value of $C_6$ denotes the dispersion coefficient of repulsive Van der Waals interaction between Rydberg-excited atoms separated by $R_{ik}$. For the sake of simplification, we neglect the effects of
atomic position and motion on the relative phase between individual atoms~\cite{PhysRevLett.107.093601}. When we assume $C_6/R_{ik}^6\gg g_{\rm m,o}\Omega_{\rm 1,2}/\Delta$ and all atoms in the ensemble are initially prepared in the ground state $|G\rangle=|g_1,g_2,..., g_{N_a}\rangle$, they will collectively transition to a symmetric single-atom excitation state due to strong vdW interactions~\cite{PhysRevA.88.033422}. Taking the simplest two-atom system as an example, the associated collective blockade radius is defined as $R_b=(\Delta C_6/\sqrt{2}g_{\rm m}\Omega_{\rm 1})^{1/6}$~\cite{PhysRevLett.87.037901}, where the interaction energy matches the excitation linewidth, which is mainly determined by the power broadening $\sqrt{2}g_{\rm m}\Omega_{\rm 1m}/\Delta$. Assuming that both atoms are driven with the same Rabi frequency, they will suppress the double-excitation state but can transition to a symmetric single-excitation state $|\psi_+\rangle=(|gr_2\rangle+|r_2g\rangle)/\sqrt{2}$. When we consider the $N_a$-scale system, the collective blockade radius is $R_b=(\Delta C_6/\sqrt{N_a}g_{\rm m}\Omega_{\rm 1})^{1/6}$, and the corresponding collective state can be described by 
\begin{equation}
|R\rangle=\frac{1}{\sqrt{N_a}}\sum_{i=1}^{N_a}|g_1,g_2,...,r_{2i},...,g_{N_a}\rangle.
\end{equation}
This means that all atoms share a single Rydberg excitation, which is the collective Rydberg state in the two-level model of superatomic states mentioned in the main text. The Hamiltonian describing the coupling of this superatom to the microwave and optical cavities is exactly expressed by Eq.~\eqref{eq3}, where the coupling strengths are scaled by $\sqrt{N_a}$. 

\bibliography{ref}

\end{document}